\documentclass[journal]{IEEEtran}
\usepackage{amsmath,epsfig}
\usepackage{paralist,amssymb, amsfonts}
\usepackage{algorithmic}
\usepackage{algorithm}
\usepackage{cite}
\usepackage{tabularx}
\usepackage{setspace}
\usepackage{multirow}
\DeclareMathOperator*{\argmin}{arg\,min}

\begin{document}

%
\title{Exploiting Prior Knowledge in Compressed Sensing Wireless ECG Systems}
%
%
%

\author{Luisa~F.~Polan\'{i}a,~\IEEEmembership{Student Member,~IEEE,}
        Rafael~E.~Carrillo,~\IEEEmembership{Member,~IEEE,}
        Manuel~Blanco-Velasco,~\IEEEmembership{Senior Member,~IEEE,}
        and~Kenneth~E.~Barner,~\IEEEmembership{Senior Member,~IEEE}
\thanks{L.F. Polan\'{i}a and K.E. Barner are with the Department
of Electrical and Computer Engineering, University of Delaware, Newark, DE, 19711 USA (e-mail: lfpolani@udel.edu; barner@udel.edu).}
\thanks{R.E. Carrillo is with the Institute of Electrical Engineering, \'{E}cole Polytechnique F\'{e}d\'{e}rale de Lausanne (EPFL), CH-1015 Laussane, Switzerland (e-mail: rafael.carrillo@epfl.ch).}
\thanks{M. Blanco-Velasco is with the Department of Signal Theory and Communications, Universidad de Alcal\'{a}, Madrid 28805, Spain (e-mail: manuel.blanco@uah.es).}}

%
%

\markboth{}%
{Shell \MakeLowercase{\textit{et al.}}: Bare Demo of IEEEtran.cls for Journals}
%



\maketitle

\begin{abstract}
Recent results in telecardiology show that compressed sensing (CS) is a promising tool to lower energy consumption in wireless body area networks for electrocardiogram (ECG) monitoring. However, the performance of current CS-based algorithms, in terms of compression rate and reconstruction quality of the ECG, still falls short of the performance attained by state-of-the-art wavelet based algorithms. In this paper, we propose to exploit the structure of the wavelet representation of the ECG signal to boost the performance of CS-based methods for compression and reconstruction of ECG signals. More precisely, we incorporate prior information about the wavelet dependencies across scales into the reconstruction algorithms and exploit the high fraction of common support of the wavelet coefficients of consecutive ECG segments. Experimental results utilizing the  MIT-BIH Arrhythmia Database show that significant performance gains, in terms of compression rate and reconstruction quality, can be obtained by the proposed algorithms compared to current CS-based methods.
\end{abstract}
\begin{IEEEkeywords}
Electrocardiogram (ECG), wireless body area networks (WBAN), compressed sensing (CS), wavelet transform.
\end{IEEEkeywords}

%
\IEEEpeerreviewmaketitle

\section{Introduction}
\IEEEPARstart{A} WIRELESS body area network (WBAN) is a radio frequency-based wireless networking technology that connects small nodes  with sensor and/or actuator capabilities in, on, or around a human body. WBANs promise to revolutionize health monitoring by allowing the transition from centralized health care services to ubiquitous and pervasive health monitoring in every-day life. One of the major challenges in the design of such systems is energy consumption, as WBANs are battery powered~\cite{Huas09}.

The  WBAN energy consumption can be divided into three main processes: sensing, wireless communication and data processing. However, the process that consumes most of the energy is the wireless transmission of data~\cite{Huas09}, which indicates that some data reduction operation should be performed at the sensor node to reduce the energy cost of the network. In addition, data reduction can supplement the bandwidth constraints of the WBAN when many sensor nodes are required to measure different physiological signals. Mamaghanian \textit{et al}.~\cite{Mama11} recently proposed to use CS to lower energy consumption and complexity in WBAN-enabled ECG monitors. Compressed sensing is an emerging field that exploits the structure of signals to acquire data at a rate proportional to the information content rather than the frequency content, therefore allowing sub-Nyquist sampling~\cite{Dono06,Cand08}. When this sampling strategy is introduced in WBANs, it gives rise to sensor nodes that efficiently acquire a small group of random linear measurements that are wirelessly transmitted to remote terminals. Indeed, sensor nodes can achieve high compression of ECG data with low computational cost when using a sparse binary sensing matrix~\cite{Mama11}.

Most state-of-the-art algorithms for ECG compression are based on wavelet transforms because of their straightforward implementation and desirable time and frequency localization properties~\cite{Iste01,Hilt97,Zhita00}. The main works in this area are characterized by hierarchical tree structures, such as embedded zero-tree wavelet (EZW)~\cite{Hilt97} and set partitioning in hierarchical tree (SPIHT)~\cite{Zhita00}, which leverage the correlations of the wavelet coefficients across scales within a hierarchically decomposed wavelet tree. Even though wavelet-based methods offer good performance in terms of compression, CS-based methods exhibit superior energy efficiency due to their lower processing complexity at the encoder~\cite{Mama11, Dixo12}.

The application of CS in WBAN-enabled ECG monitors is still at its infancy. Dixon \textit{et al}.~\cite{Dixo12} studied several design considerations for  CS-based ECG telemonitoring via a WBAN, including the encoder architecture and the design of the measurement matrix. Their results show high compression ratios using a 1-bit Bernoulli measurement matrix. Zhilin \textit{et al}.~\cite{Zhan13} introduced CS for wireless telemonitoring of non-invasive fetal ECG. They exploited the block-sparsity structure of the signals in the time domain and used block sparse Bayesian learning (BSBL) for the reconstruction. In~\cite{Mama11a}, Mamaghanian \textit{et al}. showed significant gains in compression and robustness attained by exploiting structural ECG information. In a more recent work, Mamaghanian \textit{et al}.~\cite{Mama12} proposed a new CS architecture based on random demodulation techniques for ambulatory physiological signal monitoring in embedded systems. Their results exhibit better energy efficiency than traditional acquisition systems.

Our previous works in the area include CS-based algorithms for ECG compression with a focus on algorithms achieving joint reconstruction of ECG cycles by exploiting correlation between adjacent heartbeats~\cite{Pola11,Pola12,Pola12b}. We also proposed a novel CS-based approach to recover ECG signals contaminated with electromyographic (EMG) noise using fractional lower-order moments~\cite{Pola12a}.

Even though previous works validate the potential of CS  for  energy-efficient ECG compression~\cite{Mama11,Dixo12,Zhan13}, the performance of CS in terms of compression rate and ECG reconstruction quality is still unsatisfactory when compared to the results attained by state-of-the-art algorithms based on exploiting magnitude correlation across wavelet subbands~\cite{Hilt97,Zhita00}. Most previous CS-based ECG compression works exploit only the sparsity of the signal, ignoring important signal structure information that can be known \textit{a priori} and lead to enhanced reconstruction results.

The main contribution of this paper is the application of a  CS algorithm that enables exploitation of ECG signal structure in the reconstruction process. Two key signal structure properties are incorporated into the proposed algorithm. The first property captures the wavelet domain dependencies across subbands and the second exploits the significant fraction of common wavelet coefficient support for consecutive ECG segments. The proposed algorithm falls within the framework of model-based CS~\cite{Bara10}---a new CS framework based on unions of subspaces---that can enhance signal reconstruction while reducing the number of measurements. The motivation for using a model-based approach is that it enables the incorporation of structural dependencies between the locations of the signal coefficients caused by R wave events. However, it is worth mentioning that the proposed algorithm differs from traditional model-based recovery algorithms~\cite{Bara10, Mama11a} in two respects. First, it uses prior support estimate knowledge to leverage the small variation in the support set of adjacent data sequences. Second, it excludes the selection of coefficients from the lowest-energy wavelet subband.

The performance of the proposed method is evaluated using the MIT-BIH Arrythmia Database~\cite{Gold00}. Given that most CS ECG application works use the BPDN algorithm~\cite{Mama11,Dixo12}, system-level comparisons are provided based on implementations of the BPDN and proposed reconstruction algorithms. Simulations are also performed with the bound-optimization-based BSBL algorithm, previously employed by Zhilin \textit{et al}.~\cite{Zhan13} for non-invasive fetal ECG, in order to compare with other structured sparsity-based CS reconstruction algorithms. Results indicate that the proposed algorithms outperform both BPDM and bound-optimization-based BSBL in terms of compression rate and reconstruction quality.

The organization of the paper is as follows. Section II presents a brief review of CS and model-based CS~\cite{Bara10}. Section III describes the connected subtree structure encountered in the wavelet representation of the ECG and studies the support variation across consecutive ECG segments. In Section IV, the proposed method is presented. Numerical results for the proposed method and comparisons with a benchmark state-of-the-art algorithm for ECG compression, SPIHT, are presented in Section V. Finally, we conclude in Section VI with closing remarks.


\section{Background}
\subsection{Compressed sensing}
Let $x\in\mathbb{R}^N$ be a signal that is either
$K$-sparse or compressible in an orthogonal basis $\Psi$. Thus, the signal $x$ can be approximated by a linear combination of a small number of vectors from $\Psi$, \textit{i.e.} $x\approx\sum_{i=1}^{K} s_i\psi_i$, where $K\ll N$. Let $\Phi$ be an $M\times N$ sensing matrix, $M<N$. Compressed sensing~\cite{Dono06, Cand08} addresses the recovery of $x$ from linear measurements of the form $y=\Phi x=\Phi\Psi s$. Defining $\Theta=\Phi\Psi$, the measurement vector becomes $y=\Theta s$. Compressed sensing results show that the signal $x$ can be reconstructed from $M=\mathcal{O}(K \text{log}(N/K))$ measurements if the matrix $\Theta$ satisfies some special conditions, \textit{e.g.}, the restricted isometry property~\cite{Cand08}, the mutual coherence~\cite{Bruc09}, or the null space property~\cite{Dono06}. In real scenarios with noise, the signal $s$ can be reconstructed from $y$ by solving the convex optimization problem

\begin{equation}\label{eq1}
\min_{s} \frac{1}{2}\|y-\Theta s\|_{2}^{2}+\lambda \|s\|_{1},
\end{equation}
with  $\lambda$ a regularization parameter that balances sparsity and reconstruction fidelity. The optimization problem in (\ref{eq1}) is referred to as basis pursuit denoising (BPDN)~\cite{Chen98}.

In addition to  convex optimization methods, a family of iterative greedy algorithms~\cite{Trop07,Need08,Need09} has received significant attention due to the algorithmic simplicity and low complexity. Two iterative algorithms of importance for the problem at hand are Compressive Sampling Matching Pursuit (CoSaMP)~\cite{Need08} and Iterative Hard Thresholding (IHT)~\cite{Blum09}.

\subsubsection{CoSaMP}
 CoSaMP is an iterative greedy algorithm that offers the same theoretic performance guarantees as even the best convex optimization approaches~\cite{Need08}. At each iteration, several components of the vector $s$ are selected based on the largest correlation values between the columns of $\Theta$ and the residual vector. If they are found sufficiently reliable, their indices are added to the current support estimate of $s$. This is repeated until all the recoverable portion of the signal is found.
 \subsubsection{Iterative Hard Thresholding}
IHT is a powerful method for sparse recovery that converges to a local minimum of the problem statement
\begin{equation}\label{eqIHT}
\min_{s} \|y-\Theta s\|_{2}^{2} \text{ subject to } \|s\|_{0}\leq K,
\end{equation}
by using the recursion
 \begin{equation}\label{BPD}
s_{i+1}=H_{K}(s_{i}+\Theta^T(y-\Theta s_{i}))
\end{equation}
where $H_{K}$ is a non-linear operator that sets to zero all elements other than the $K$ largest elements (in magnitude).
IHT is a very simple algorithm that recovers sparse and compressible vectors with a minimal number of observations and with near optimal accuracy, whenever the matrix $\Theta$ has a small restricted isometry constant~\cite{Blum09}.
 \subsection{Model-based Compressed Sensing}
Model-based compressive sensing is a new paradigm that aims to capture the inter-dependency structure in the support of the large signal coefficients using a union-of-subspaces model~\cite{Bara10}. This model decreases the degrees of freedom of the signal by allowing only some configurations of support for the largest coefficients.

Let $x$ be a $K$-sparse signal. Then, $x$ lies in $\Sigma_K\subset \mathbb{R}^N$, which is a union of $\binom{N}{K}$ subspaces of dimension $K$. A union-of-subspaces model allows only some $K$-dimensional subspaces in $\Sigma_K$ and leads to representations that incorporate signal structure. Let $x|_\Omega$ denote the coordinates of $x$ corresponding to the set of indices $\Omega \subseteq {1, \ldots, N}$, and let $\Omega^C$ represent the complement of $\Omega$. Then, the signal model $ \mathcal{M}_K$ is defined as the union of $m_K$ canonical $K$-dimensional subspaces

\begin{equation}
    \mathcal{M}_K=\bigcup_{m=1}^{m_K} \mathcal{X}_m,    \mathcal{X}_m=\{x: x|_{\Omega_m} \in \mathbb{R}^K, x|_{\Omega_m^C}=0\}
\end{equation}
where $\{\Omega_1, \ldots, \Omega_{m_K}\}$ is the set of all allowed supports with $|\Omega_{m}|=K$ for each $m=1\ldots m_K$. It should be noted that $\mathcal{M}_K\subseteq \Sigma_k$ and that $\mathcal{M}_K$ contains $m_K\leq \binom{N}{K}$ subspaces. Signals from $\mathcal{M}_K$ are called $K$-model sparse.

A similar treatment is applied to compressible signals. A compressible signal $x \in R^N$ that is nearly $K$-model sparse can be approximated to the best model-based approximation in $\mathcal{M}_K$. The $\ell_2$ error produced by the approximation is given by
\begin{equation}
    \sigma_{\mathcal{M}_K}=\inf_{\bar{x}\in\mathcal{M}_K} \|x-\bar{x}\|_2.
\end{equation}
The algorithm that provides the best $K$-term approximation of the signal $x$ under the model $\mathcal{M}_K$ is denoted as $\mathbb{M}(x,K)$. Thus, the error $\sigma_{\mathcal{M}_K}$ can also be written as $\sigma_{\mathcal{M}_K}= \|x-\mathbb{M}(x,K)\|_2$.
A sparsity model $\mathcal{M}=\{\mathcal{M}_1, \mathcal{M}_2, \ldots\}$ produces nested approximations if the support of $\mathbb{M}(x,K^\ast)$ contains the support of $\mathbb{M}(x,K)$ for all $K<K^\ast$. If the signal model produces nested approximations, then the support of the difference vector $\mathbb{M}(x,jK)-\mathbb{M}(x,(j-1)K)$ lies in a small union of subspaces. These difference vectors form sets of residual subspaces. The $j^{th}$ set of residual subspaces of size $K$ is defined as
\begin{equation}
\begin{split}
\mathcal{R}_{j,K}(\mathcal{M})=\{u\in\mathbb{R}^N \text{ s.t. } u=\mathbb{M}(x,jK)-\mathbb{M}(x,(j-1)K)\\
\text{ for some $x\in\mathbb{R}^N$}\},
\end{split}
\end{equation}
for $j=1, \ldots, \lceil N/K\rceil.$
A structured compressible signal $x$ can be robustly recovered from the compressive measurements $y=\Phi x$ if the matrix $\Phi$ satisfies the restricted amplification property (RAmP)~\cite{Bara10}. A matrix $\Phi$ has the $(\epsilon_K,r)$-restricted amplification property for the residual subspaces $R_{j,K}$ of model $\mathcal{M}$ if
\begin{equation}
\|\Phi u\|_2^2\leq(1+\epsilon_K)j^{2r}\|u\|_2^2
\end{equation}
for any $u \in R_{j,K}$ and for each $1\leq j \leq \lceil N/K\rceil$.

Baraniuk \textit{et al.}~\cite{Bara10} incorporated the union-of-subspaces models into two well-known CS recovery algorithms, CoSaMP and IHT, through a single modification in the algorithms. The modification, in practice, replaces the best $K$-term approximation with a best $K$-term model-based approximation.

\subsection{Tree-structured sparsity models}
One example of a structured sparsity model is that encountered in signals whose most significant wavelet coefficients are organized into a tree structure, and where the largest coefficients cluster along the branches of the tree~\cite{Bara10}.

Consider an $N$-dimensional signal $x$. Given a wavelet function $\psi$ and a scaling function $\varphi$, the wavelet representation of $x$ is defined in terms of shifted versions of $\varphi$ and shifted and dilated versions of $\psi$
\begin{equation}
x=\displaystyle\sum\limits_{i=0}^{N_L-1} a_{L,i}\varphi_{L,i}+\displaystyle\sum\limits_{j=1}^{L}\displaystyle\sum\limits_{i=0}^{N_j-1}d_{j,i}\psi_{j,i},
\end{equation}
where $j$ denotes the scale of analysis and $L$ denotes the coarsest scale. $N_j=N/2^j$ indicates the number of coefficients at scale $j\in\{1,\ldots,L\}$ and $i$ represents the position, $0\leq i\leq N_j-1$. The wavelet transform consists of the scaling coefficients $a_{L,i}$ and wavelet coefficients $d_{j,i}$. Using the previous notation, we write $x=\Psi s$, where $\Psi$ is the orthogonal matrix containing the wavelet and scaling functions as columns and $s=[d_{1,0} \ldots d_{1,N_1-1} \ldots d_{L,0} \ldots d_{L,N_L-1} a_{L,1} \ldots a_{L,N_L-1}]^T$ is the vector of scaling and wavelet coefficients. The vector $s$ can be decomposed into $L+1$ subvectors. The first $L$ subvectors are denoted by $d_j$, $j=1,\ldots,L$, and the $j$th subvector contains all of the wavelet coefficients for scale $j$. The last subvector corresponds to the scaling coefficients and is denoted as $a_L$. Thus, $s$ can also be written as $s=[d_1 d_2 \ldots d_L a_L]^T$.

The wavelet atoms form a binary tree structure where the wavelet coefficient $d_{j,i}$ is the parent of its two children $d_{j+1,2i}$ and $d_{j+1, 2i+1}$. This nesting property causes rapid transitions and other singularities to manifest as chains of large coefficients along the branches of the wavelet tree~\cite{Crou98}. This gives rise to the concept of a connected subtree, which refers to a connected set of nodes $\Omega$ meeting the condition that whenever a coefficient $d_{j,i}\in\Omega$, then its parent also belongs to $\Omega$. In~\cite{Bara10}, Baraniuk \textit{et al.} assumed $N_L=1$ for simplicity and defined the structured sparsity model $\mathcal{T}_K$ as the union of all $K$-dimensional subspaces corresponding to supports $\Omega$ that form connected subtrees,
\begin{multline}
    \mathcal{T}_K=\bigg\{x=a_{L,0}\varphi_{L,0}+\displaystyle\sum\limits_{j=1}^{L}\displaystyle\sum\limits_{i=0}^{N_j-1}d_{j,i}\psi_{j,i}: d|_{\Omega^C}=0, \\
    |\Omega|=K, \Omega \mbox{ forms a connected subtree} \bigg\},
\end{multline}
where $\Omega^C$ denotes the complement of the set $\Omega$.
To find the best $K$-term tree-based approximation,  Baraniuk \textit{et al.} used the condensing sort and select algorithm (CSSA)~\cite{Bara02}, which solves for
\begin{equation}\label{BPD2}
x^{\ast}=\argmin_{\bar{x}\in \mathcal{T}_k} \|x-\bar{x}\|_2
\end{equation}
by using a two-stage process. The first stage merges the non-monotonic segments of the tree branches with an iterative sort-and-average routine. The second stage simply sorts the wavelet coefficients once they are organized in a monotonic non-increasing sequence along the branches out from the root. The principle behind the CSSA algorithm is that the standard $K$-term approximation coincides with the subtree approximation when the wavelet data is monotonically non-increasing along the tree branches.


\section{MOTIVATION}
In this section, we analyze the structure of the wavelet representation of ECG signals to motivate the incorporation of prior information in CS-based recovery algorithms for ECG reconstruction. We concentrate on exploiting two key properties of the ECG wavelet coefficients. The first property is the connected subtree structure formed by the largest (in magnitude) wavelet coefficients, and the second property is the high fraction of common support between adjacent ECG segments.
\label{ssec:CSPKS}
\begin{figure}[t]
\centering{ 
\includegraphics[width = \columnwidth]{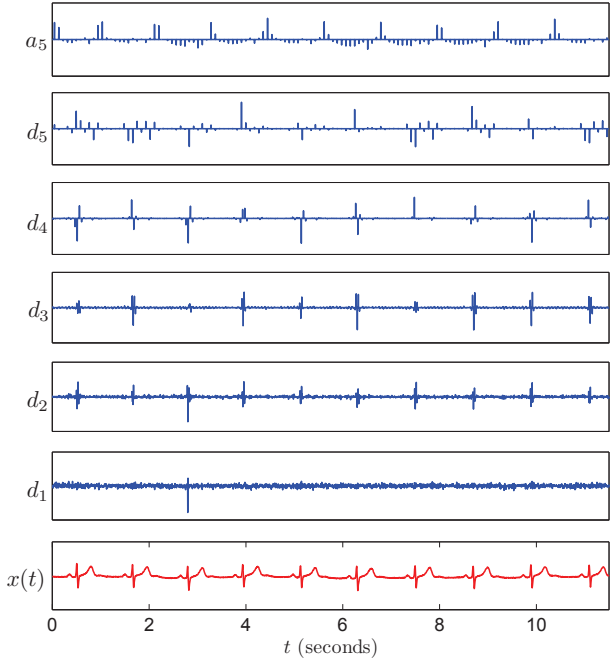}}
\caption{Daubechies-4 coefficients subvectors $d_j$, $j=1\ldots5$, and $a_5$ for ECG time series. } \label{fig:stack}
\end{figure}

\subsection{Connected subtree structure of ECG wavelet coefficients}
Sharp transition regions in the time domain generate large magnitude wavelet coefficients that persist along the branches of the wavelet tree, forming connected rooted subtrees~\cite{Crou98}. This behavior is also present in the wavelet representation of ECG signals and is connected with R wave events. This idea is illustrated in Fig.~\ref{fig:stack}, which shows that coefficients subvectors $d_{j}$, $j=1,\ldots,5$, and $a_5$, corresponding to the Daubechies-4 wavelet transform of an ECG time series using a decomposition level $L=5$. The subvectors are plotted as rows stacked on top of each other. The ECG time series, located at the bottom of Fig.~\ref{fig:stack}, corresponds to an extract of 11.5 seconds from record 117 of the MIT-BIH Arrhythmia Database. Each coefficient vector is time-shifted so that the tree structure can be clearly identified. For visualization purposes, the magnitude of the coefficients is normalized so that the Euclidean norm of each wavelet subband is unity.

Examining the wavelet representation in Fig.~\ref{fig:stack}, it is noticed that the large coefficients are aligned and propagate across scales, forming a connected tree structure. This persistence property is mainly noticed in subbands $d_4, d_3$, and $d_2$. These results suggest that the tree-structured sparsity model described in Section II-C is an appropriate model to represent the support of the most significant ECG wavelet coefficients. The tree structure is intrinsically related to the shape of the ECG cycles. When compared with the ECG time series, it is noted that the large coefficients are connected with the QRS complexes, which can be regarded as sharp transition regions of the signal. We propose to exploit the tree-structured sparse representation as additional prior information for the CS reconstruction of ECG signals. As stated by the model-based CS framework~\cite{Bara10}, knowledge of the signal support potentially leads to high quality reconstruction using fewer measurements, and thus higher compression performance.
\begin{figure}[t]
\centering{ 
\includegraphics[width = \columnwidth]{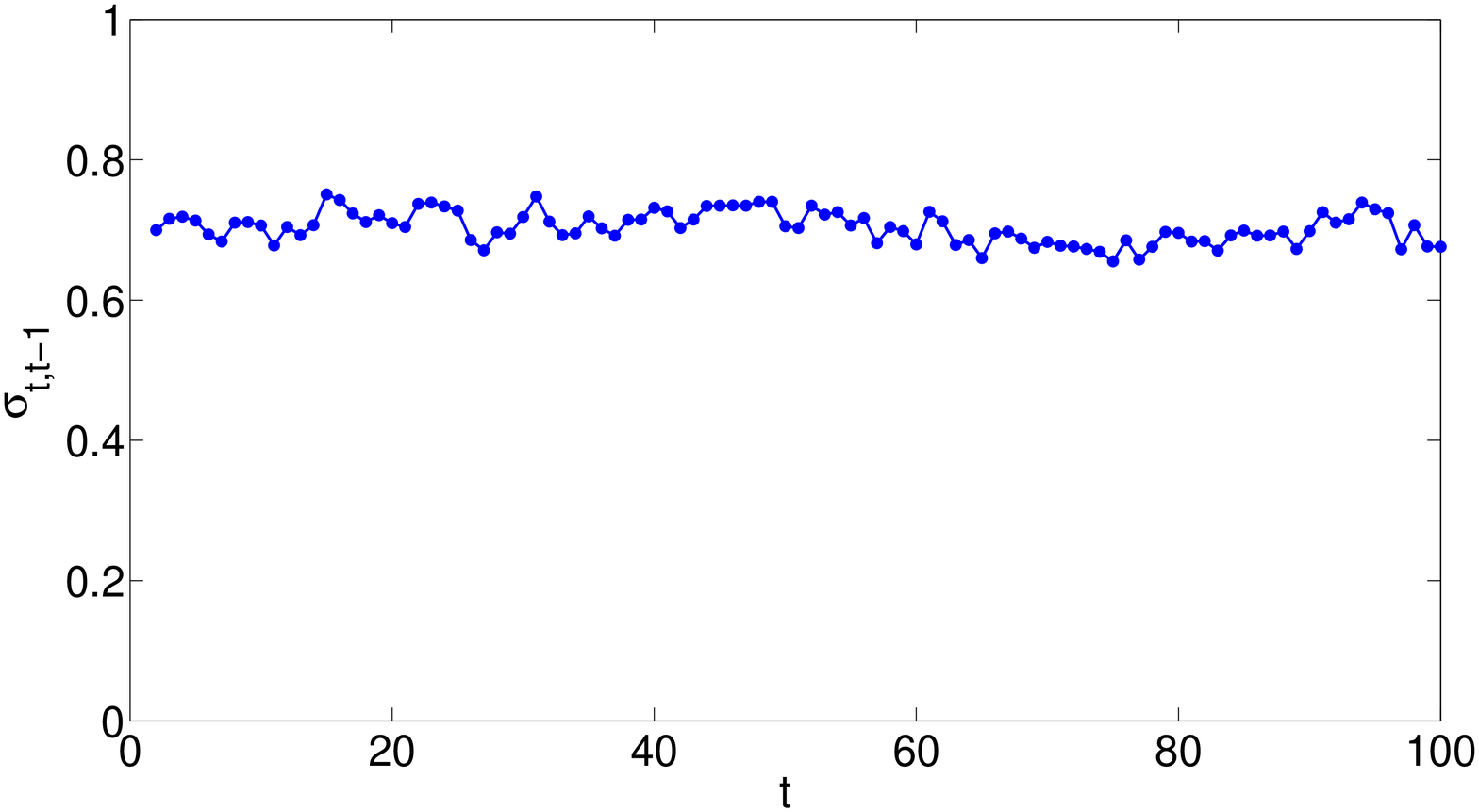}}
\caption{Fraction of common support between consecutive ECG segments. The results are averaged over the selected data set from the MIT-BIH Arrhythmia Database.} \label{fig:2}
\end{figure}

\subsection{Common support between consecutive ECG segments}
\label{ssec:CSPKSS}
In this section, we study how the support of two consecutive ECG segments varies. Consider the 10-min long single leads extracted from records 100, 101, 102, 103, 107, 109, 111, 115, 117, 118 and 119 in the MIT-BIH Arrhythmia Database, which are sampled at 360 Hz, and form 100 consecutive sequences of $2048$ samples for each record. Let $\Omega_t$  denote the support of the $K=225$ largest (in magnitude) wavelet coefficients of sequence $s_t$, where $t=1,\ldots,100$ denotes the order of the sequences. Consider the fraction of common support between two consecutive sequences, denoted as $\sigma_{t,t-1}$, which is defined as
\begin{equation}\label{LIP100}
\sigma_{t,t-1}=\frac{|\Omega_t\cap \Omega_{t-1}|}{|\Omega_t|}, ~~t=2, \ldots, 100.~~
\end{equation}

\begin{figure*}[t]
\centering{ 
\includegraphics[width = \textwidth]{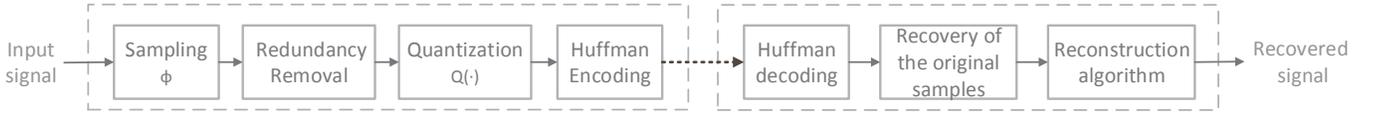}}
\caption{Block Diagram of the proposed ECG compression method.} \label{fig:3}
\end{figure*}

The averaged results over the entire ECG data set are illustrated in Fig.~\ref{fig:2}, where the data points are connected for visualization purposes only. The fraction of common support is high, always greater than 0.65, which indicates that we can use the support of the previous data sequence to improve the estimate of the support of the current data sequence. In designing the CS-based reconstruction algorithm, we propose to include support information of previous reconstructed data sequences to improve the performance. A related work by Wang \textit{et al.}~\cite{Wang10} shows how to iteratively enhance recovery of a signal by solving basis pursuit in the first iteration to obtain a support estimate, solving the problem of basis pursuit with partially known support~\cite{Vasw09} with this support estimate, and repeating the steps. Their work attaines exact reconstruction guarantees for a single iteration even if the initial support estimate includes a small fraction of indices outside the true support.

\section{METHODS}
\label{ssec:CSPKSSS}
The proposed ECG compression method based on compressive sensing for potential implementation in WBAN-based telemonitoring systems is described in this section. The method is summarized in Fig.~\ref{fig:3}. The process initiates with the sampling of the ECG signals through linear random measurements, followed by a redundancy removal module and a quantization stage. The quantized samples are entropy coded and transmitted to a remote terminal where the reconstruction is performed. The presented contribution focuses on the reconstruction algorithm that relies on prior knowledge of ECG wavelet coefficient structure to reduce the required number of measurements and improve reconstruction quality. A more detailed explanation of each stage is given below.

\subsection{Sampling and Encoding}
The first step addresses the sampling of consecutive ECG segments of length $N$. Let $x$ denote an ECG segment. The information we acquire about $x$ can be described by $y=\Phi x$, where $\Phi$ is a $M \times N$ matrix. In order to recover the best $K$-term model-based approximation of the original signal, the matrix $\Phi$ needs to satisfy the restricted amplification property (RAmP). It is known that sub-Gaussian matrices meet this condition with high probability~\cite{Bara10}. Here we build the entries of the matrix $\Phi$ by independently sampling from a symmetric Bernoulli distribution (P($\Phi_{i,j}=\pm1/\sqrt{M}=1/2$)) in order to facilitate an efficient hardware implementation. The use of Bernoulli matrices, as compared to other sub-Gaussian matrices, results in simpler circuit complexity, data storage, and computation requirements~\cite{Chen12}.

To realize the analog hardware implementation of CS matrices, Mamaghanian \textit{et al}.~\cite{Mama12} recently proposed the spread spectrum random modulator pre-integrator. This architecture starts with a pre-modulation block, followed by a random demodulation pre-integrator architecture, which is composed of parallel channels of random demodulators. Each random demodulator is, in turn, composed of three stages. The first stage refers to the multiplication of the input signal with a pseudo-random sequence that takes values $\pm1$ with equal probability. Such sequences are generated with a pseudo-random bit sequence generator. The second stage incorporates a low-pass filter to avoid aliasing and the final stage corresponds to a standard ADC. Their implementation achieves $43\%$ energy saving, compared to traditional Nyquist sampling, when applied to EEG and ECG signals.

The use of the described sampling procedure results in similar adjacent measurement vectors. This is caused by the quasi-periodic nature of the ECG signal and the use of a fixed sampling matrix. To further improve the compression and reduce the amount of redundant information, we implemented the same redundancy removal and entropy coding stages proposed by Mamaghanian \textit{et al}.~\cite{Mama11}. The redundancy removal stage computes the difference between two consecutive measurement vectors and only transmits this difference to the quantization module. In~\cite{Mama11}, it was shown that the variance of the difference signals between consecutive measurement vectors is lower than the variance of the original measurement vectors, which leads to a reduction in the number of bits for signal representation, from 12 to 9 bits. It is worth mentioning that the experiments in~\cite{Mama11} were applied to non-aligned ECG segments, which implies that the QRS complexes were located at different locations and, nevertheless, the corresponding measurements exhibited significant redundancy to be removed. An 8-bit optimal scalar quantizer designed with the Lloyd-Max algorithm is utilized~\cite{Gers92} and entropy coding stage uses Huffman coding to further increase the compression ratio.

\subsection{Reconstruction algorithm using prior information}
The original samples must be recovered from the transmitted difference between consecutive vectors. Therefore, at the remote terminal, Huffman decoding followed by recovery of the original samples is employed. For reconstruction of the ECG signal, two iterative algorithms that are easily modified to incorporate prior ECG wavelet representation structure information are proposed. The algorithms are CoSaMP and IHT, which were previously modified by Baraniuk \textit{et al.}~\cite{Bara10} to incorporate structured sparsity models. Their modification results in replacing the nonlinear sparse approximation step with a structured sparse approximation. These algorithms are known in the literature as model-based CoSaMP and model-based IHT; both have provable robust guarantees. One of the main properties of model-based CoSAMP is that robust signal recovery requires only a number of measurements that is proportional to the sparsity level of the signal.

\begin{figure}[t]
\centering{ 
\includegraphics[width = \columnwidth]{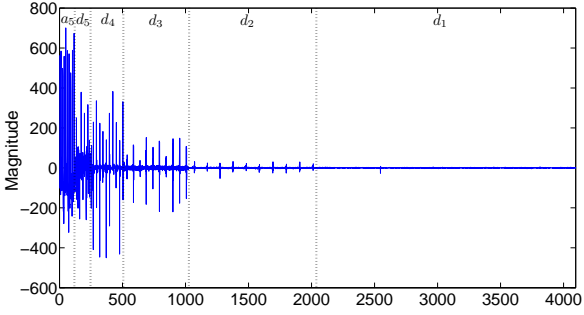}}
\caption{Wavelet decomposition of level $L=5$ for ECG time series using Daubechies-4.} \label{fig:4}
\end{figure}

The sparsity model employed in this paper corresponds to a modified version of the tree-structured sparsity model described in section Section II-C. The modification is based on the fact that ECG wavelet coefficients at scale $j=1$ correspond to coefficients with nearly zero magnitude, as illustrated in Fig.~\ref{fig:4}, and should  therefore not be included in the best $K$-term model-based signal approximation. Thus, we select $N_L$ scaling coefficients and define $\mathcal{T}_K^\star$ as
\begin{multline}
    \mathcal{T}_K^\star=\bigg\{x=\displaystyle\sum\limits_{i=0}^{N_L-1} a_{L,i}\varphi_{L,i}+\displaystyle\sum\limits_{j=1}^{L}\displaystyle\sum\limits_{i=0}^{N_j-1}d_{j,i}\psi_{j,i}: \\ \text{supp}(d_{1,i})\in\Omega^C \text{ for all } i, d|_{\Omega^C}=0, |\Omega|=K,\\ \Omega \mbox{ forms a connected subtree} \bigg\}.
\end{multline}

Denote the algorithm that finds the best $K$-term tree-based approximation by $\mathbb{T}(x,K)$. That is,
\begin{equation}\label{BPD10}
\mathbb{T}(x,K)=\argmin_{\bar{x}\in \mathcal{T}_k^\star} \|x-\bar{x}\|_2.
\end{equation}
The model-based CoSaMP and the model-based IHT, using the tree-structured sparsity model $ \mathcal{T}_K^\star$, are summarized in Algorithm 1 and 2, respectively, and are referred to as modified model-based CoSaMP (MMB-CoSaMP) and modified model-based IHT (MMB-IHT). The halting criterion for both algorithms can be a fixed number of iterations or a bound on the residual norm, $\|r_j\|_2\leq \epsilon$, for some predetermined $\epsilon>0$. Note that in the algorithm tables, the Moore-Penrose pseudo-inverse of $\Theta$ is denoted by $\Theta^\dagger$, $\Theta_T$ denotes the submatrix obtained by extracting the columns of $\Theta$ corresponding to the indexes in $T$, $b|_{T}$ represents the entries of $b$ corresponding to the set of indices $T$, and the support of $s$ is denoted by $\text{supp}(s)$. Also, we use the condensing sort and select algorithm~\cite{Bara02} to solve for the optimization problem in \eqref{BPD10}.

As in model-based CoSaMP, MMB-CoSaMP starts the iteration by calculating the correlation values between the columns of $\Theta$ and the residual of the previous iteration $r_{j-1}$ (step 4).  The correlation values are used to find support of the best $2K$-term tree-based approximation (step 5), which is  subsequently merged with the support of the previous iteration (step 6). The next step refers to solving a least-squares problem to approximate the target signal on the updated support (step 7). To enforce a sparse solution, the algorithm finds the best $K$-term tree-based approximation of the least squares solution (step 8) and finalizes with the update of the residual (step 9).

Similarly, MMM-IHT also resembles model-based IHT. The algorithm iteratively solves (\ref{eqIHT}) by moving in the opposite direction of the gradient of $\|y-\Theta s\|_{2}^{2}$ at each iteration (step 4). MMM-IHT enforces a sparse solution by selecting the best $K$-term tree-based approximation of $b$ (step 5) and finalizes with an update of the residual (step 6). It is worth emphasizing that the proposed algorithms exploit the wavelet tree structure in the steps that refer to the best $K$-term tree-based approximation defined in \eqref{BPD10}. Both MMB-IHT and MMB-CoSaMP differ from the traditional model-based approaches in the initialization step and in the $K$-term tree-based approximation of the signal.

Unlike the traditional model-based approaches, the proposed algorithms always include the scaling coefficients in the $K$-term tree-based approximation of the signal, which is of great importance given that the scaling coefficients accumulate most of the ECG signal energy. Indeed, in a previous work~\cite{Carr10}, we show that enforcing CS greedy algorithms to select the atoms corresponding to the scaling coefficients can lead to a significant reduction in the number of measurements while improving the accuracy of the ECG reconstruction.

The initialization step is a modification with respect to traditional approaches because it incorporates support information from the previously reconstructed signal. In section III-B, it is noted that a high fraction of support is shared between two consecutive ECG segments. The two consecutive ECG segments in the wavelet domain are denoted by $s^t$ and $s^{t-1}$. With the aim of exploiting this information, the first signal estimate is determined by solving a least squares problem using the support of the previously reconstructed ECG segment, $\Lambda=\text{supp}(s^{t-1})$. The contribution of this estimate is subtracted from the measurement vector and iterated on the residual. These steps are implemented in the initialization of Algorithm 1 and 2. In both algorithms, MMB-IHT and MMB-CoSaMP, the support set is refined through iterations by addition of promising new atoms and deletion of unnecessary atoms. In this way, we expect the algorithms to preserve the common support ($\text{supp}(s^{t}) \cap \text{supp}(s^{t-1})$) and to replace the unnecessary atoms ($\text{supp}(s^{t-1})\setminus\text{supp}(s^{t})$) with the innovation support of the current ECG sequence, \textit{i.e.} ($\text{supp}(s^{t})\setminus\text{supp}(s^{t-1})$).

\section{EXPERIMENTAL RESULTS}
To validate the proposed methods, compressed measurements are generated using a set of records from the MIT-BIH Arrhythmia Database~\cite{Gold00} as the original signals. Every file in the database contains two lead recordings sampled at 360 Hz with 11 bits per sample of resolution. However, since body area networks adopt lower sampling frequencies than 360 Hz, each ECG recording is resampled at 250 Hz. The sampling frequency of 250 Hz is commonly used for ECG monitoring in body area networks~\cite{Pant10}. Experiments are carried out and averaged over 10-min long single leads extracted from records 100, 101, 102, 103, 107, 109, 111, 115, 117, 118 and 119. This data set was proposed in~\cite{Zhita00}; it consists of a variety of signals with different rhythms, wave morphologies and abnormal heartbeats. Results are presented for averages of 100 repetitions of each experiment, with a different realization of the random measurement matrix at each time.
\begin{algorithm}[t]
\caption{MMB-CoSaMP}\label{alg:alg1}
\begin{algorithmic}[1]
\REQUIRE Matrices $\Theta=\Phi\Psi$ and $\Psi$, measurements $y$, sparsity level $K$, support of previously reconstructed ECG segment $\Lambda$, structured sparse approximation algorithm $\mathbb{T}$.
\STATE Initialize $\hat{s}_0|_{\Lambda}=\Theta_{\Lambda}^\dagger y$, $\hat{s}_0|_{\Lambda^C}=0$, $r_0=y-\Theta\hat{s}_0$, $j=0$.
\WHILE{halting criterion false}
\STATE $j\leftarrow j+1$
\STATE $e\leftarrow \Theta^T r_{j-1}$
\STATE $\Omega \leftarrow \text{supp}(\Psi^T\mathbb{T}(\Psi e,2K))$
\STATE $T \leftarrow \Omega\cup \text{supp} (\hat{s}_{j-1})$
\STATE $b|_{T} \leftarrow \Theta_{T}^\dagger y $, $b|_{T^C} \leftarrow 0$
\STATE $\hat{s}_{j} \leftarrow \Psi^T\mathbb{T}(\Psi b,K)$
\STATE $r_j \leftarrow y-\Theta\hat{s}_j$
\ENDWHILE
\RETURN $\hat{x} \leftarrow \Psi\hat{s}_j$
\end{algorithmic}
\end{algorithm}

\subsection{Performance Evaluation}
The compression ratio (CR), the percentage root-mean-square difference (PRD), the normalized version of PRD (PRDN), the quality score (QS), and the the reconstruction SNR (R-SNR) are used as performance measures. The CR is defined as the number of bits required for the original signal over the number of bits required for the compressed signal. Here the original signals refer to the resampled ECG records at 250 Hz. The reconstruction SNR is defined as
\begin{equation}\label{BPD1807}
\text{R-SNR}=10\text{log}_{10}\frac{\|x\|_2^2}{\|x-\hat{x}\|_2^2},
\end{equation}
where $x$ and $\hat{x}$ denote the $N$-dimensional original and reconstructed signals, respectively. The PRD is defined as $\text{PRD}=(\|x-\hat{x}\|_2/\|x\|_2)\times100$. Let $e$ denote an $N$-dimensional vector of ones and $\bar{x}$ be the mean value of $x$. The PRDN is defined as $\text{PRDN}=(\|x-\hat{x}\|_2/\|x-\bar{x}e\|_2)\times100$, and the QS as $\text{QS}=\text{CR}/\text{PRD}$~\cite{Fira08}.

\subsection{Practical considerations}
The length of the ECG segments is set to $N=256$, so that the acquisition time can be sufficiently short (approximately 1 sec at the rate of 250 Hz) for real-time monitoring. The orthogonal Daubechies-4 wavelets is chosen as the sparsifying transform. For the reconstruction algorithms, the halting criterion is either a maximum number of iterations (we selected 70 for our simulations) or a bound on the residual norm, $\|r_t\|_2\leq \epsilon$. We selected $\epsilon=1\times10^{-3}\|y\|_2$.

\begin{figure}[t]
\centering{ 
\includegraphics[width = \columnwidth]{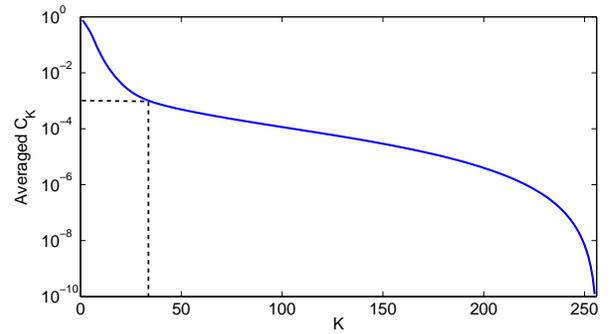}}
\caption{Sequence of residual energy averaged over the selected set of records from the MIT-BIH Arrhythmia Database.} \label{fig:5}
\end{figure}

For the sparsity level, a sequence of residual energy is defined. The elements in the vector of wavelet coefficients $s$ are ordered according to their square magnitudes, such that
\begin{equation}\label{BPD180687}
|s_{(1)}|^2 \geq |s_{(2)}|^2 \geq \ldots \geq |s_{(N-1)}|^2 \geq |s_{(N)}|^2.
\end{equation}
The sequence of residual energy is defined as
\begin{equation}\label{BPD180688}
C_K=\frac{\sum\limits_{j=1}^{N} |s_{(j)}|^2-\sum\limits_{j=1}^{K} |s_{(j)}|^2}{\sum\limits_{j=1}^{N} |s_{(j)}|^2},~~K=1, \ldots, N.~~
\end{equation}
From each record of the data set, 300 ECG segments of length $N=256$ are selected. The sequence of residual energy is averaged over all the selected ECG segments and over all the different records. The results are plotted in~Fig.~\ref{fig:5} in logarithmic scale. The sparsity level is selected as the value of $K$ that satisfies averaged $C_K=0.001$, which corresponds to $K=34$ and is indicated in Fig.~\ref{fig:5} with dotted lines. This result indicates that the most significant 34 wavelet coefficients approximately accumulate 99.999\% of the total signal energy.
\label{ssec:CSPKSSSS}
\begin{algorithm}[t]
\caption{MMB-IHT}\label{alg:alg2}
\begin{algorithmic}[1]
\REQUIRE  Matrices $\Psi$, $\Phi$, and $\Theta=\Phi\Psi$, measurements $y$, sparsity level $K$, support of previously reconstructed ECG segment $\Lambda$, structured sparse approximation algorithm $\mathbb{T}$.
\STATE Initialize $\hat{s}_0|_{\Lambda}=\Theta_{\Lambda}^\dagger y$, $\hat{s}_0|_{\Lambda^C}=0$, $\hat{x}_0=\Psi \hat{s_0}$, $r_0=y-\Phi\hat{x}_0$, $j=0$
\WHILE{halting criterion false}
\STATE $j\leftarrow j+1$
\STATE $b\leftarrow \hat{x}_{j-1}+\Phi^T r_{j-1}$
\STATE $\hat{x}_j\leftarrow\mathbb{T}(b,K)$
\STATE $r_j\leftarrow y-\Phi\hat{x}_j$
\ENDWHILE
\RETURN $\hat{x} \leftarrow \hat{x}_j$
\end{algorithmic}
\end{algorithm}

\begin{table*}[t]
\footnotesize
\renewcommand{\arraystretch}{1.3}
\caption{PRD obtained by the proposed algorithms for different sensing matrices}
\label{table_example}
\centering
\begin{tabular*}{0.8\textwidth}{@{\extracolsep\fill}c|c|cccccccccc}
\hline
\multirow{2}{1.5cm}{Reconstruction algorithm}&\multirow{2}{1.5cm}{Sampling matrix}&\multicolumn{10}{c}{CR}\\
\cline{3-12}
&&3.5&4&4.5&5&5.5&6&6.5&7&7.5&8\\
\hline
\multirow{3}{1.5cm}{MMB-IHT}& Bernoulli  &3.31&3.32&3.32&3.33&3.35&3.55&3.81&4.05&4.43&4.91\\
&Matrix I &  3.42&3.42&3.47&3.7&3.85&4.23&4.8& 5.71&6.7&7.79\\
&Matrix II & 3.42&3.43&3.45&3.55&3.67&3.99&4.32&4.68&5.2 & 5.82\\
\hline
\multirow{3}{1.5cm}{MMB-CoSaMP}&Bernoulli & 2.98&2.99&2.99&3.11&3.24&3.49&4.02&4.8&9.74&25.58\\
&Matrix I &  2.99&2.99&3.04&3.12&3.62&3.92&5.12&6.9&14.01&28.15\\
&Matrix II & 3.01&3.01&3.03&3.07&3.4&3.7&4.8&6.2& 12.9&27.74\\
\hline
\end{tabular*}
\end{table*}

\begin{figure}[t]
\centering{ 
\includegraphics[width = \columnwidth]{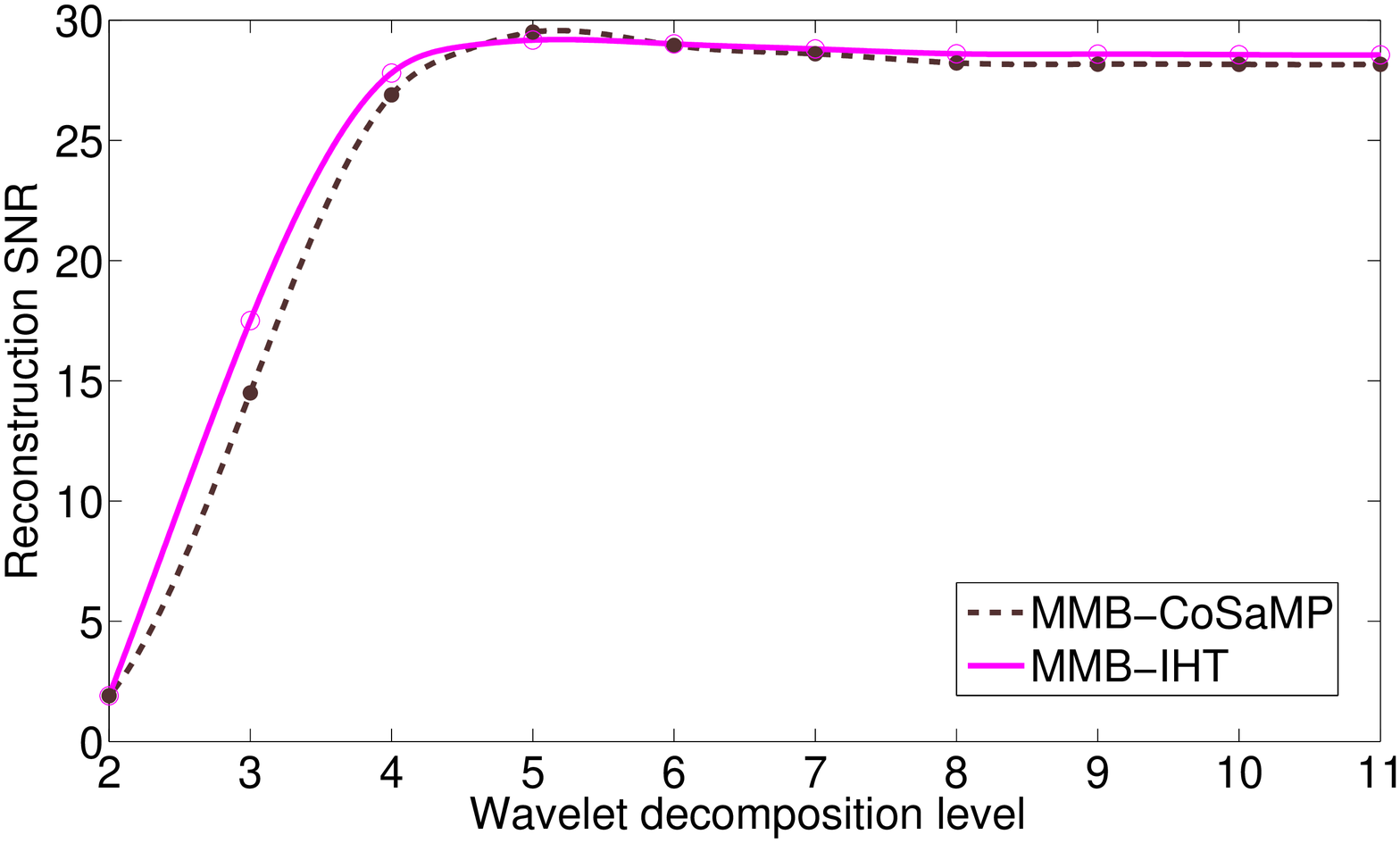}}
\caption{Reconstruction SNR as a function of the wavelet decomposition level. Number of measurements $m=3K$.} \label{fig:dl}
\end{figure}

Given that the ECG recordings are sampled at 250Hz, the sampling interval becomes $\Delta t=1/250$ seconds. A decomposition level $L=5$ is utilized so that the scaling coefficients vector $a_5$, associated with a physical scale of $2^5\Delta t=2^5/250=0.13 \text{ seconds}$, approximately isolate the T waves, and thus the detail coefficients capture the QRS complex. In this way, the largest (in magnitude) detail coefficients are expected to exhibit a connected subtree structure caused by the QRS complexes, as shown in Fig.~\ref{fig:stack}. The ECG reconstruction quality, using both MMB-CoSaMP and MMB-IHT algorithms, is evaluated as a function of the wavelet decomposition level. The results are averaged over the selected set of records for a number of measurements $m=3K$. The results in Fig.~\ref{fig:dl} indicate that the selection of $L=5$ is indeed a good choice for the decomposition level. The data points are connected for visualization purposes only.


Sparse binary matrices have recently been proposed for CS ECG since they lead to fast computations and low-memory requirements~\cite{Mama11}. An experiment is designed to test the performance of the proposed algorithms when using sparse binary matrices. Similarly to the work in~\cite{Mama11}, two types of sparse binary matrices, matrix I and matrix II, are employed. Matrix I has only $q\ll N$ nonzero elements in each column. Each nonzero element takes the value $1/\sqrt{q}$ and its location is chosen randomly. Matrix II has only $q\ll N$ nonzero elements in each column. Each nonzero element takes the value $\pm1/\sqrt{q}$ with equal probability and its location is chosen randomly. In~\cite{Mama11}, the value of $q=\lfloor0.025\times N\rfloor$ provides a good tradeoff between execution time and reconstruction quality. The same relation between the value of $q$ and $N$ is selected for the proposed experiment, $q=\lfloor0.025\times N\rfloor=6$. The reconstruction performance of the proposed algorithms using the Bernoulli matrix, described in Section IV.A, and the sparse binary matrices is shown in Table I. The results correspond to averaged PRD values over the entire set of selected records. Even though all the matrices have a similar performance for $\text{CR}\leq6$, the performance of the sparse binary matrices deteriorates for $\text{CR}>6$. Hence, the dense Bernoulli matrix is selected as the sampling matrix for the proposed methods. Nevertheless, the minimal detriment to performance suggests that sparse binary sensing matrices are a plausible alternative for CS ECG.

\subsection{Evaluation of ECG reconstruction using the proposed method}
This section presents an experiment to evaluate how the reconstruction of ECG signals changes as the amount of prior knowledge of the support set varies. The best performance is achieved with the oracle estimate. Assume an oracle reveals the support set $\Omega$ of the $K$ most significant wavelet coefficients of the signal of interest. The oracle estimate corresponds to the least squares projection onto the subspace spanned by the columns of $\Theta$ with indices in $\Omega$. This experiment also considers the case where no prior information is known about the support set $\Omega$, and the signal is reconstructed using the traditional CoSaMP~\cite{Need08} and the traditional IHT~\cite{Blum09}. The performance of all these methods is compared with the performance attained by the proposed algorithms, MMB-CoSaMP and MMB-IHT.

\begin{figure}[t]
\centering{ 
\includegraphics[width = \columnwidth]{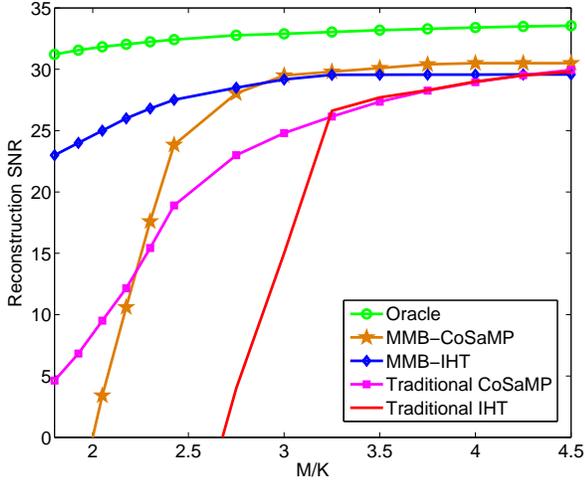}}
\caption{Comparison of MMB-CoSaMP and MMB-IHT with CoSaMP, IHT, and the oracle estimate. Reconstruction SNR averaged over all the records of the selected data set for different number of measurements.} \label{fig:202}
\end{figure}

The results are averaged over the set of selected records and illustrated in Fig.~\ref{fig:202}. Given that the objective of this first experiment is only to evaluate the reconstruction of the proposed scheme, we restrict our method to the sampling and reconstruction of the signals and exclude the redundancy removal, quantization, and entropy coding stages. For this experiment, the reconstruction SNR is used to evaluate the quality of the recovered signals as a function of the oversampling ratio $M/K$.

As shown in Fig.~\ref{fig:202}, the proposed algorithms outperform the traditional CoSaMP and IHT, indicating that exploiting the connected subtree structure of the most significant wavelet coefficients, as well as the common support between consecutive ECG segments, results in a large reduction of the required number of measurements to achieve successful recovery of ECG signals. It is also noted from Fig.~\ref{fig:202} that the algorithms based on IHT require fewer measurements than the algorithms based on CoSaMP to attain good reconstruction. The results of the proposed methods are the closest to the best achievable performance obtained by the oracle estimate.

\subsection{Evaluation of compression performance}
\begin{figure}[t]
\centering{ 
\includegraphics[width = \columnwidth]{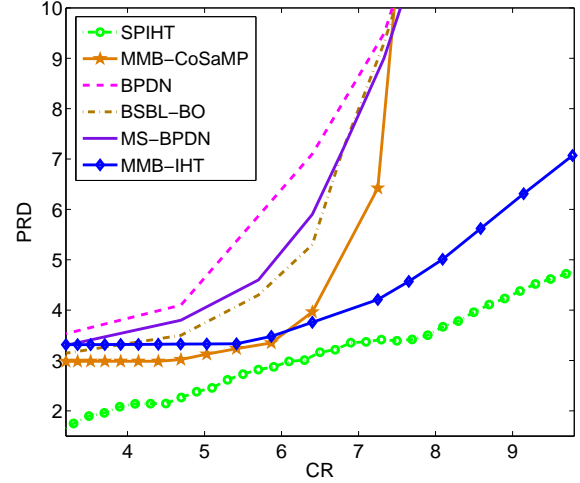}}
\caption{Compression performance evaluation of the compression scheme using the proposed algorithms (MMB-CoSaMP, MMB-IHT), the BSBL-BO, the MS-BPDN, and the BPDN algorithms. The results of SPIHT are included as a baseline for comparisons.} \label{fig:203}
\end{figure}

For the second experiment, the reconstruction algorithm in the compression scheme, Fig.~\ref{fig:3}, is varied. The proposed MMB-CoSaMP and  MMB-IHT are first used as reconstruction algorithms, and their results are compared with BPDN, the bound-optimization-based BSBL algorithm, and an overcomplete dictionary-based reconstruction algorithm. The results are also compared with SPIHT, a state-of-the-art algorithm for ECG compression. The same entropy coding stage of the proposed method is added to SPIHT to ensure fairness in the comparison. The results of the PRD as a function of the compression ratio are illustrated in Fig.~\ref{fig:203}. The results are averaged over the entire set of selected records.

Basis pursuit denoising is the reconstruction algorithm selected by Mamaghanian \textit{et al}.~\cite{Mama11} and Dixon \textit{et al}.~\cite{Dixo12} for the recovery of ECG signals. However, the results in Fig.~\ref{fig:203} indicate that the proposed compression scheme works significantly better when MMB-IHT and MMB-CoSaMP are used as reconstruction algorithms than when BPDN is employed. These results are expected, as the proposed reconstruction algorithms exploit prior knowledge of the signal structure, unlike BPDN, which only leverages the sparsity of the signals.

Unlike BPDN, the bound-optimization-based BSBL algorithm, denoted as BSBL-BO, provides flexibility to exploit the block structure and  intra-block correlation of the signal sparsity pattern. Even though it was previously employed to reconstruct non-invasive fetal ECG~\cite{Zhan13}, it is also successfully applied in the recovery of adult ECG in the wavelet domain. The reason is the clustering property of the ECG wavelet coefficients that suggests the use of a block-sparsity model. This property refers to the tendency of the large coefficients to cluster together into blocks, as it is suggested by Fig.~\ref{fig:stack}. For the implementation of the BSBL-BO algorithm, the block partition was set to $h=15$ and the maximum number of iterations to 30. Even though the BSBL-BO algorithm offers performance superior to BPDN, it is outperformed by MMB-IHT and MMB-CoSaMP. This result suggests that the connected subtree structured sparsity model may be more appropriate to represent the largest (in magnitude) ECG wavelet coefficients than is the block sparsity model. In addition, the incorporation of prior support knowledge also contributes to the superior performance attained by the proposed methods.

It is of interest to compare the performance of the proposed algorithms with overcomplete dictionary-based reconstruction methods. In a previous work~\cite{Pola13}, we propose the use of a multi-scale dictionary $D\in\mathbb{R}^{N\times J}$, $J>N$, for CS ECG, with the aim of combining the advantages of multi-scale representations using wavelets with the benefits of dictionary learning. The dictionary $D$ is divided into subdictionaries according to the corresponding wavelet subband and each subdictionary is learned separately. The idea behind this approach is to exploit correlations within each wavelet subband. The multi-scale dictionary-based algorithm, denoted as MS-BPDN, aims to solve problem (\ref{eq1}) with $\Theta=\Phi\Psi D$, instead of $\Theta=\Phi\Psi$, by using basis pursuit denoising. According to the results in Fig. 8, the proposed algorithms outperform MS-BPDN because they exploit additional signal structure information. However, it is also noticed that MS-BPDN outperforms the traditional wavelet based-BPDN algorithm, which suggests the promising application of adaptive overcomplete dictionaries to CS ECG. The design of more efficient overcomplete dictionary-based reconstruction algorithms, that exploit dependencies among dictionary atoms, is left for future work.

\begin{figure*}[t]
    \centering
    \begin{tabular}{cc}
    \epsfig{figure=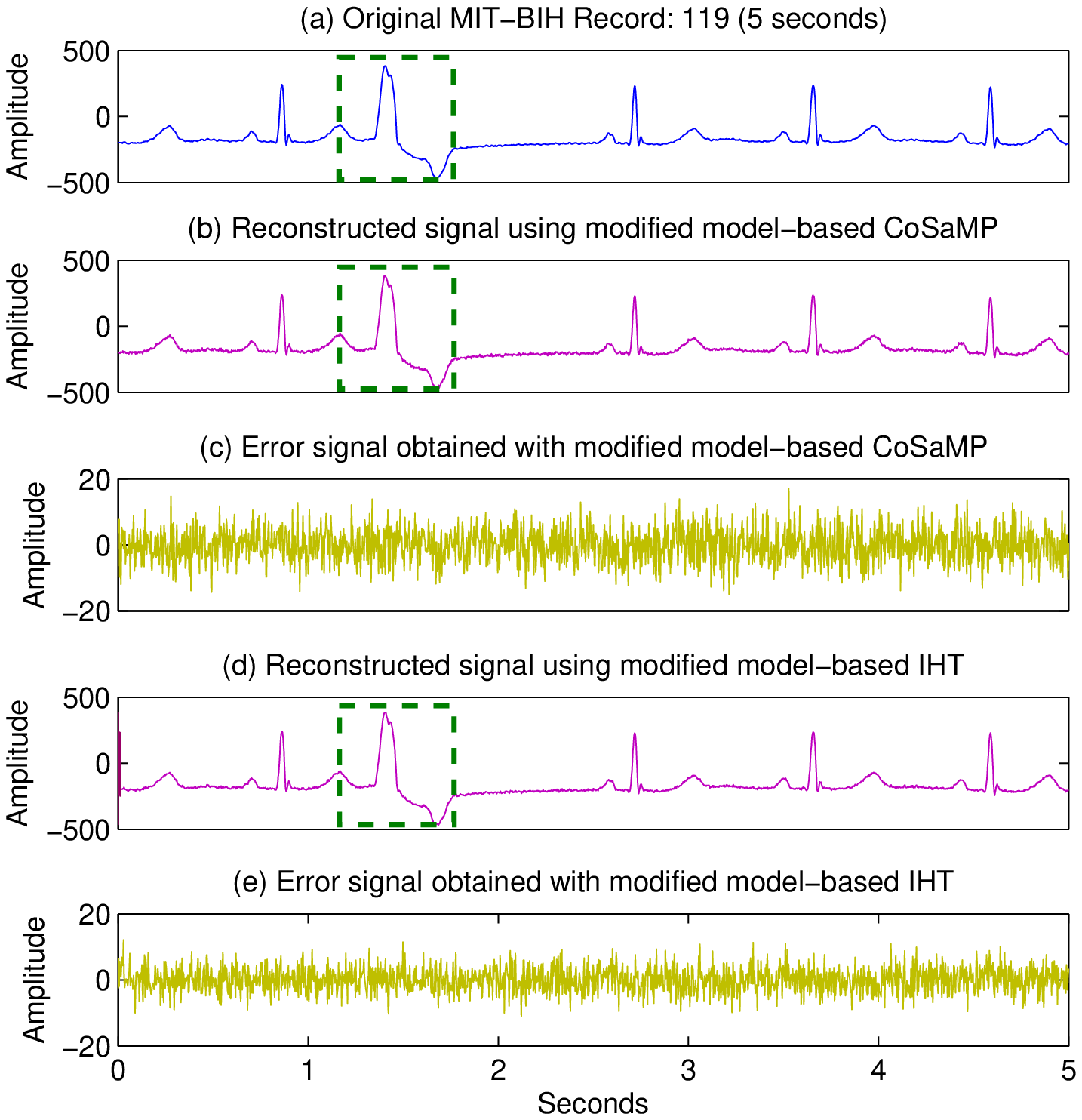,width=8.8cm}&
    \epsfig{figure=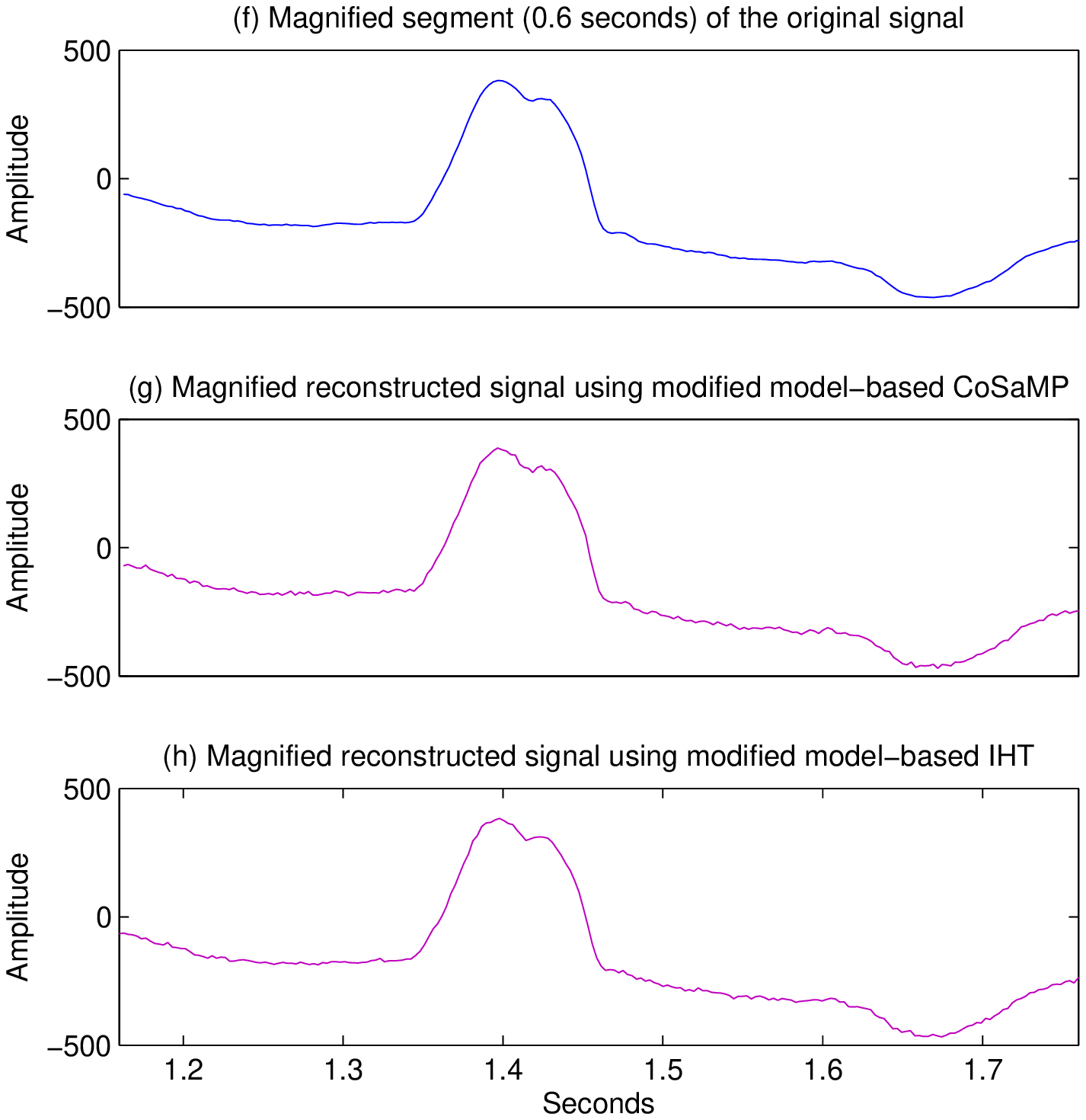,width=8.8cm}
    \end{tabular}
\caption{Visual evaluation of the reconstruction of record 119 using MMB-CoSaMP and MMB-IHT. $CR=6.4$. PRD=2.61 (MMB-CoSaMP), PRD=2.29 (MMB-IHT). Left: Recovery results for a 5-sec sequence. Right: Magnified views of the dashed boxes located on the left.}
\label{fig:visual1}
\end{figure*}
These results validate the potential of CS-based methods in achieving high compression rates while offering important advantages, as described in previous works in the area~\cite{Mama11, Dixo12}, such as higher energy and hardware efficiency at the encoder.

The results of SPIHT are illustrated in Fig.~\ref{fig:203}, with the aim of evaluating how the proposed compression scheme compares with current ECG compression methods. SPIHT is a benchmark state-of-the-art algorithm for ECG compression. SPIHT utilizes an optimized embedded coding of the wavelet coefficients that increases the compression rates and makes it outperform the results of the proposed methods, at the expense of requiring more computations. The time complexity of the SPIHT encoder is $\mathcal{O}(N \text{log}N)$~\cite{Park03}, while the matrix-vector multiplication of the CS encoder, $\Phi x$, requires only $\mathcal{O}(N)$ operations when $\Phi$ is a random symmetric Bernoulli matrix~\cite{Libe08}. Therefore, the use of a CS encoder offers a substantial complexity reduction.

In CS-based methods, there is no access to the rich structure of the wavelet coefficients at the encoder, but instead, there is access to the random measurements, which contain fewer structure properties to be exploited at the coding stage. This difference in compression performance between wavelet transform-based methods and CS-based methods was previously noted by Mamaghanian \textit{et al}.~\cite{Mama11}, who emphasized the substantial lower power consumption offered by CS-based methods. Even though the proposed algorithm does not outperform SPIHT in terms of compression ratio, it offers better compression and reconstruction performance than previously proposed CS-based methods for ECG compression, such as BPDN and BSBL-BO.

There is significant variability among the set of ECG records, and therefore, it is instructive to calculate the reconstruction performance of the proposed algorithms for each ECG recording. The results are illustrated in Table II. For almost all the records, MMB-IHT provides better performance than MMB-CoSaMP.

\begin{table}[!t]
\footnotesize
\renewcommand{\arraystretch}{1.3}
\caption{Performance of the proposed algorithms for CR=6.4}
\label{table_example}
\centering
\begin{tabular}{c|ccc|ccc}
\hline
\multirow{2}{1.5cm}{Record}&\multicolumn{3}{c|}{MMB-IHT}&\multicolumn{3}{c}{MMB-CoSaMP}\\
\cline{2-7}
&PRD&PRDN&QS&PRD&PRDN&QS\\
\hline
100&3.65&7.73&1.75&3.86&8.18&1.66\\
101&5.86&7.79&1.09&5.83&7.76&1.09\\
102&4.84&8.34&1.32&5.21&8.98&1.22\\
103&3.94&4.8&1.62&4.17&5.08&1.53\\
107&3.56&3.68&1.79&3.99&4.13&1.6\\
109&3.91&4.51&1.63&4.13&4.75&1.54\\
111&5.91&6.99&1.08&6.21&7.35&1.02\\
115&2.74&6.62&2.33&2.57&6.19&2.49\\
117&2.11&7.84&3.03&2.11&7.86&3.02\\
118&2.54&4.61&2.51&2.86&5.2&2.23\\
119&2.29&4.5&2.78&2.61&5.12&2.45\\
\hline
\end{tabular}
\end{table}

\subsection{Visual evaluation}
\begin{figure*}[t]
    \centering
    \begin{tabular}{cc}
    \epsfig{figure=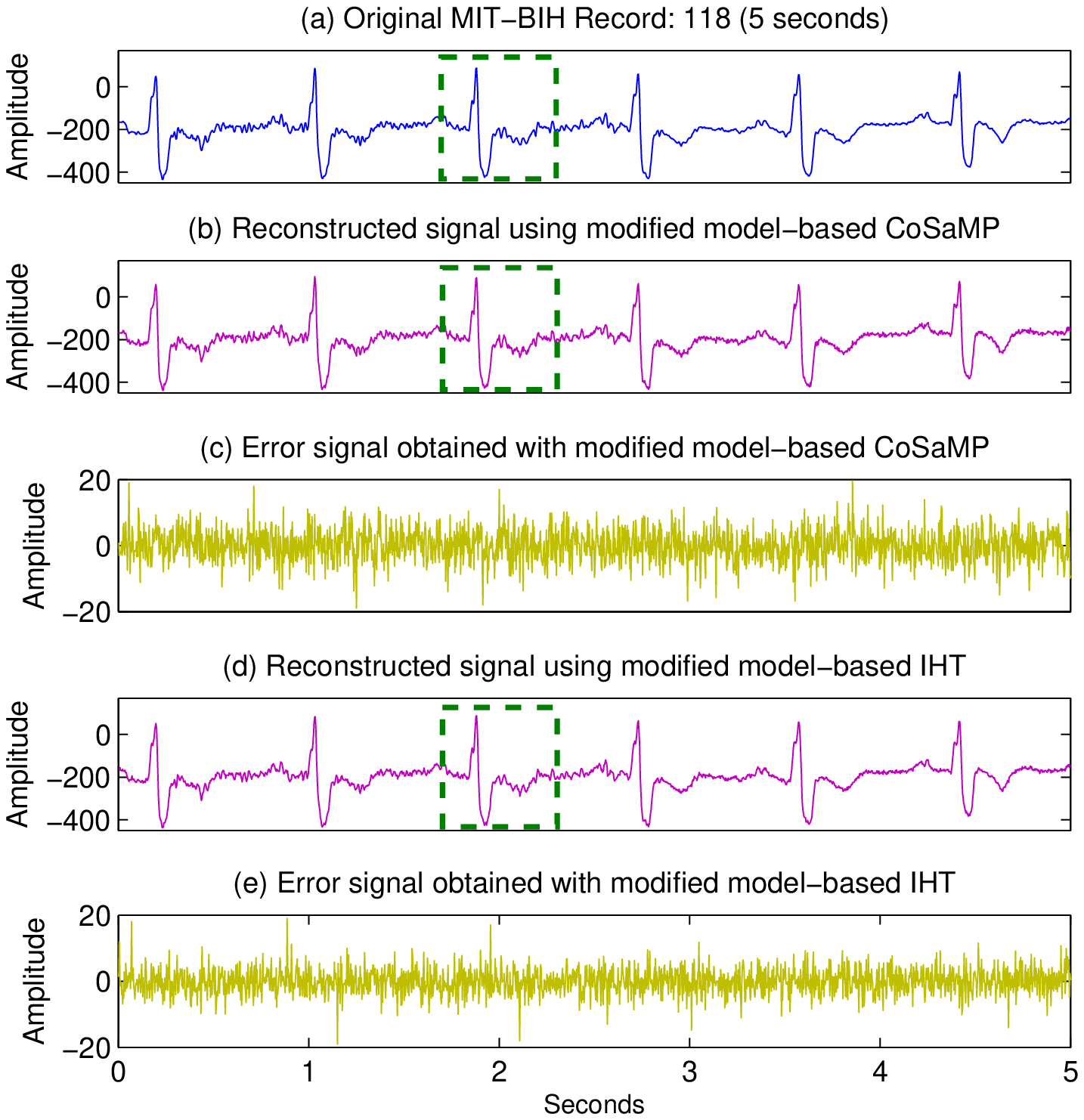,width=8.8cm}&
    \epsfig{figure=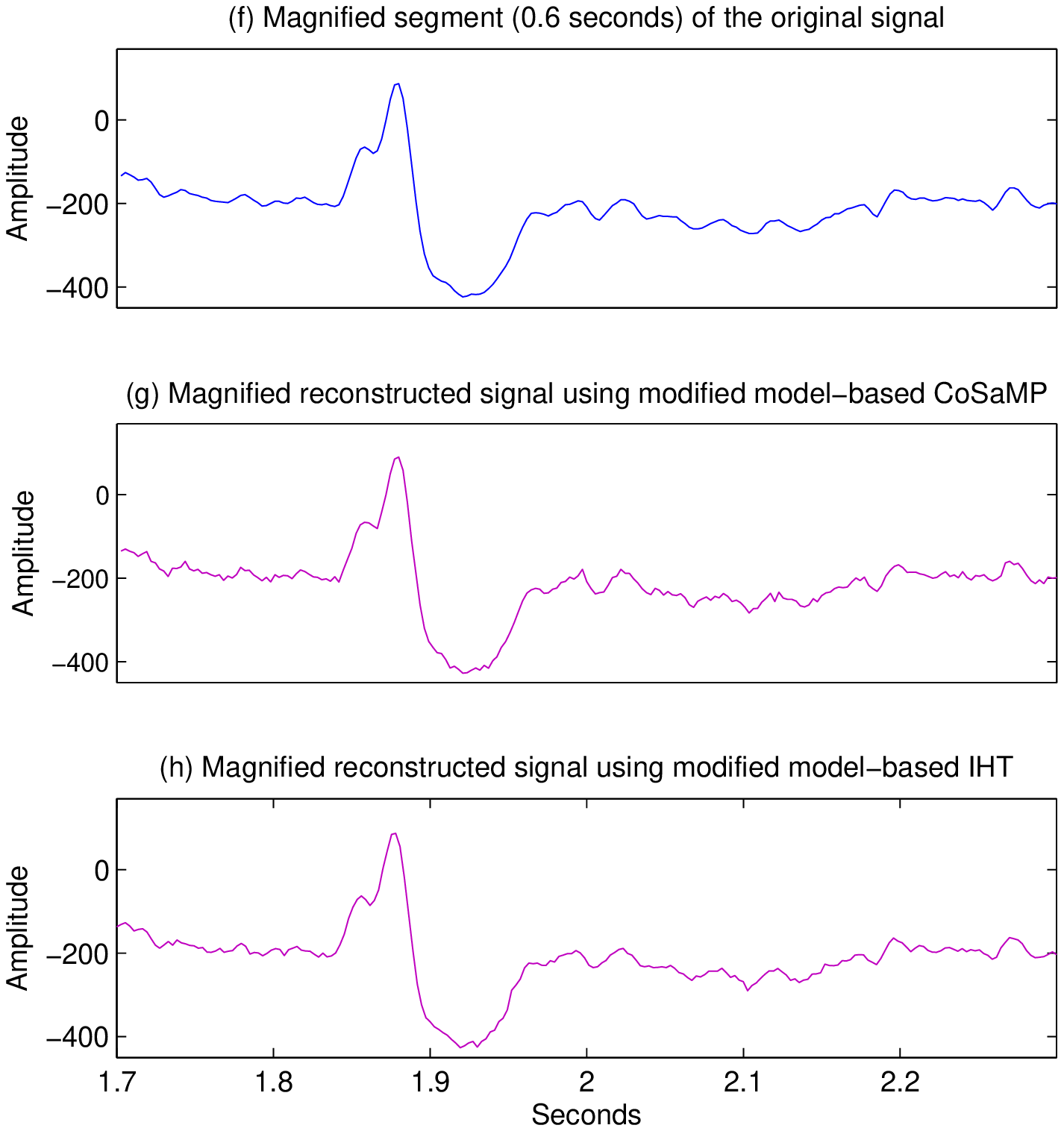,width=8.8cm}
    \end{tabular}
\caption{Visual evaluation of the reconstruction of record 118 using MMB-CoSaMP and MMB-IHT. $CR=6.4$. PRD=2.86 (MMB-CoSaMP), PRD=2.54 (MMB-IHT). Left: Recovery results for a 5-sec sequence. Right: Magnified views of the dashed boxes located on the left.}
\label{fig:visual2}
\end{figure*}

Finally, visual study of the reconstructed and error signals using the proposed compression scheme is also presented. Two records with different clinical characteristics are selected for the study: 118 and 119. Record 119 contains ventricular ectopic heartbeats while the record 118 contains right bundle branch block heartbeats. The  results for a $\text{CR}=6.4$ are shown in Fig.~\ref{fig:visual1} for record 119 and in Fig.~\ref{fig:visual2} for record 118. Magnified views of the original and reconstructed signals are included in the figures. For the two records, the recovered signals are a good estimate of the original signals and they preserve detailed information for clinical diagnosis. It should be noted that the reconstructed signals using MMB-IHT exhibit less artifacts than the reconstructed signals using MMB-CoSaMP.

\section{Conclusions}
The universal applicability of compressed sensing to sparse signals lies in the fact that no specific prior information about the signals is assumed, apart from the sparsity condition. However, for a particular application, some prior information about the signals is typically available. In this paper, we showed that for the specific application of compressed sensing to ECG compression, the appropriate incorporation of prior information into the reconstruction procedures leads to more accurate reconstruction and higher compression rates. More precisely, we exploit prior information on the connected subtree structure formed by largest (in magnitude) wavelet coefficients and the common support of the wavelet representation of consecutive ECG segments are exploited. The model-based CoSaMP and model-based IHT algorithms are modified to incorporate support knowledge from the previously reconstructed data sequence. The tree-structured sparsity model is also modified to exclude the selection of atoms from the lowest-energy wavelet subband. We justified the application of a model-based compressed sensing approach with the fact that R wave events cause a connected subtree structure of large magnitude wavelet coefficients. In addition, we also selected an appropriate wavelet decomposition level to enable the formation of such structure.

The proposed scheme was evaluated for the compression of a set of eleven ECG records from the MIT-BIH Arrhythmia Database encompassing a variety of signals with different rhythms, wave morphologies and abnormal heartbeats. The experimental results were evaluated in terms of PRD and compression ratio. The proposed method outperformed the results of previously proposed CS-based methods for ECG compression while still maintaining the low-complexity and energy-efficient implementation inherent to CS-based approaches.

\bibliographystyle{IEEEbib}
\bibliography{RGCD2}

\begin{IEEEbiography}[{\includegraphics[width=1in,height=1.25in,clip,keepaspectratio]{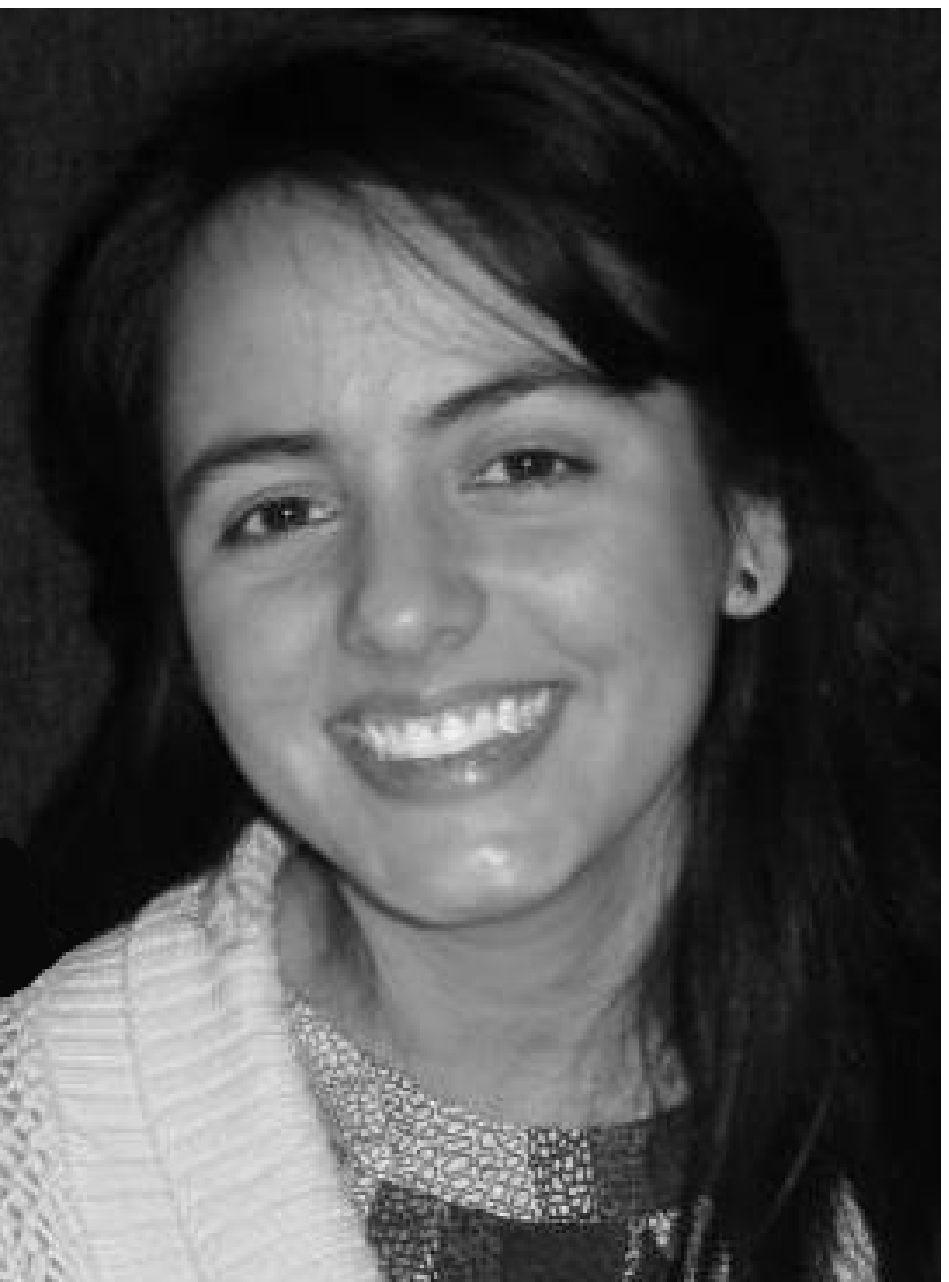}}]{Luisa F. Polan\'{i}a}
(S'12) received the B.S.E.E. degree (with honors) from the National University of Colombia, Bogot\'{a}, Colombia, in 2009. She is currently pursuing the Ph.D. degree in the Department of Electrical and Computer Engineering, University of Delaware, Newark.Her research interests include signal and image processing, compressive sensing, low-dimensional modeling and biomedical signal processing. Miss Polan\'{i}a was the recipient of the University
of Delaware Graduate Student Fellowship in 2013.
\end{IEEEbiography}

\begin{IEEEbiography}[{\includegraphics[width=1in,height=1.25in,clip,keepaspectratio]{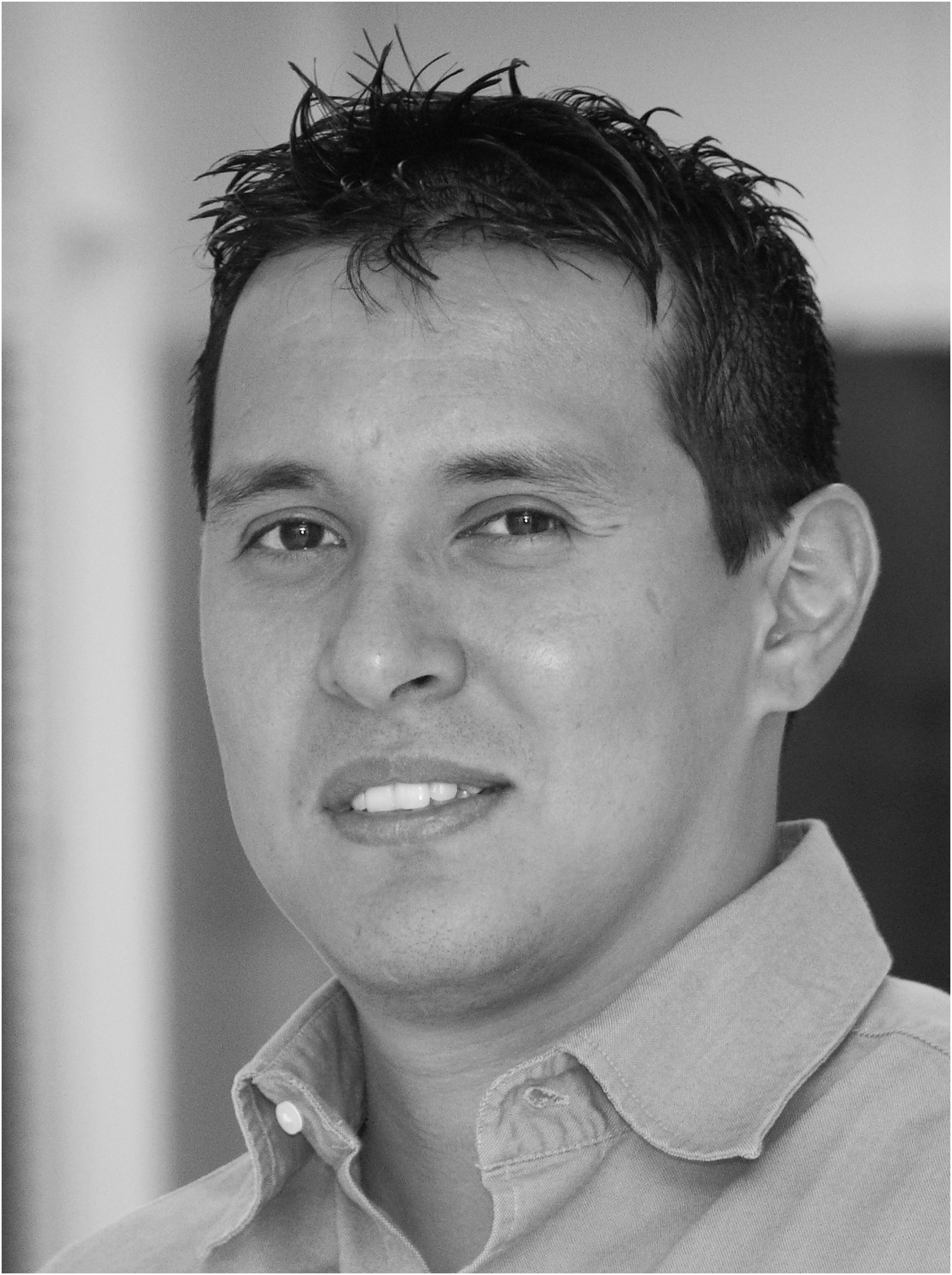}}]{Rafael E. Carrillo}
(S'07-M'12) received the B.S. and the M.S. degrees (with honors) in Electronics Engineering from the Pontificia Universidad Javeriana, Bogot\'{a}, Colombia, in 2003 and 2006 respectively and the Ph.D. degree in Electrical Engineering from the University of Delaware, Newark, DE, in 2011. He was a lecturer from 2003 to 2006 at the Pontificia Universidad Javeriana and a research assistant at the University of Delaware from 2006 to 2011. Since 2011 he is a postdoctoral researcher at the Institute of Electrical Engineering, Ecole Polytechnique F{\'e}d{\'e}rale de Lausanne, Lausanne, Switzerland.
His research interests include signal and image processing, compressive sensing, inverse problems, and robust, nonlinear and statistical signal processing. Dr. Carrillo was the recipient of the ``Mejor trabajo de grado'' award, given to outstanding master thesis at the Pontificia Universidad Javeriana, in 2006, the University of Delaware Graduate Student Fellowship in 2007 and the Signal Processing and Communications Graduate Faculty Award from the University of Delaware in 2010.
\end{IEEEbiography}
\vspace{-0.5cm}
\begin{IEEEbiography}[{\includegraphics[width=1in,height=1.25in,clip,keepaspectratio]{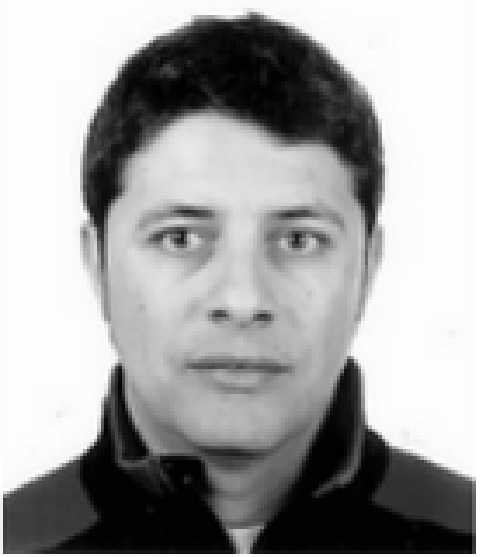}}]{Manuel Blanco-Velasco}
(S'00-M'05-SM'10) received the engineering degree from the Universidad de Alcal\'{a}, Madrid, Spain in 1990, the MSc in communications engineering from the Universidad Polit\'{e}cnica de Madrid, Spain, in 1999, and the PhD degree from the Universidad de Alcal\'{a} in 2004. From 1992 to 2002, he was with the Circuits and Systems Department at the Universidad Polit\'{e}cnica de Madrid as Assistant Professor. In April 2002, he joined the Signal Theory and Communications Department of the Universidad de Alcal\'{a} where he is now working as Associate Professor. His main research interests are biomedical signal processing and communications.
\end{IEEEbiography}
\vspace{-0.5cm}
\begin{IEEEbiography}[{\includegraphics[width=1in,height=1.25in,clip,keepaspectratio]{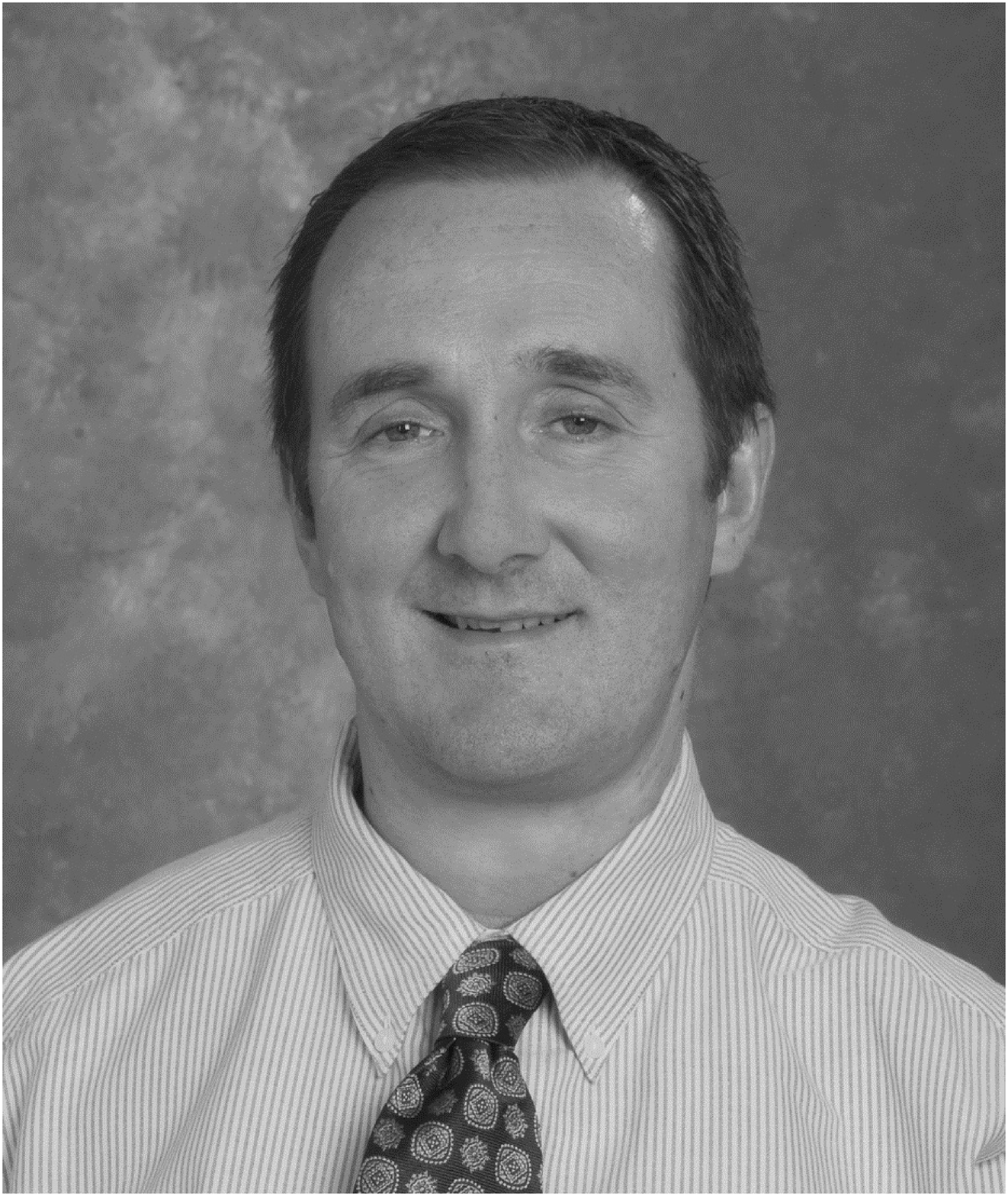}}]{Kenneth E. Barner}
(S'84-M'92-SM'00) received the B.S.E.E. degree (magna cum laude) from Lehigh University, Bethlehem, Pennsylvania, in 1987. He received the M.S.E.E. and Ph.D. degrees from the University of Delaware, Newark, Delaware, in 1989 and 1992, respectively.

He was the duPont Teaching Fellow and a Visiting Lecturer with the University of Delaware in 1991 and 1992, respectively. From 1993 to 1997, he was an Assistant Research Professor with the Department of Electrical and Computer Engineering, University of Delaware, and a Research Engineer with the duPont Hospital for Children. He is currently Professor and Chairman with the Department of Electrical and Computer Engineering, University of Delaware. He is coeditor of the book \emph{Nonlinear Signal and Image Processing: Theory, Methods, and Applications} (Boca Raton, FL: CRC), 2004. His research interests include signal and image processing, robust signal processing, nonlinear systems, sensor networks and consensus systems, compressive sensing, human-computer interaction, haptic and tactile methods, and universal access.

Dr. Barner is the recipient of a 1999 NSF CAREER award. He was the Co-Chair of the 2001 \emph{IEEE-EURASIP Nonlinear Signal and Image Processing (NSIP) Workshop} and a Guest Editor for a Special Issue of the \emph{EURASIP Journal of Applied Signal Processing} on Nonlinear Signal and Image Processing. He is a member of the Nonlinear Signal and Image Processing Board. He was the Technical Program Co-Chair for ICASSP 2005 and and previously served on the IEEE Signal Processing Theory and Methods (SPTM) Technical Committee and the IEEE Bio-Imaging and Signal Processing (BISP) Technical Committee. He is currently a member of the IEEE Delaware Bay Section Executive Committee. He has served as an Associate Editor of the \emph{IEEE Transactions on Signal Processing}, the \emph{IEEE Transaction on Neural Systems and Rehabilitation Engineering}, the \emph{IEEE Signal Processing Magazine}, the \emph{IEEE Signal Processing Letters}, and the \emph{EURASIP Journal of Applied Signal Processing}. He was the Founding Editor-in-Chief of the journal \emph{Advances in Human-Computer Interaction}. For his dissertation ``Permutation Filters: A Group Theoretic Class of Non-Linear Filters,'' Dr. Barner received the \emph{Allan P. Colburn Prize in Mathematical Sciences and Engineering} for the most outstanding doctoral dissertation in the engineering and mathematical disciplines. He is a member of Tau Beta Pi, Eta Kappa Nu, and Phi Sigma Kappa.
\end{IEEEbiography}

%










\end{document}